\documentclass[%
 aip,
 amsmath,amssymb,
 reprint,%
]{revtex4-2}

\usepackage{bm}
\usepackage{bbm}
\usepackage{dsfont}
\usepackage{graphicx}
\usepackage{mathtools}
\usepackage{array}
\usepackage{tikz}
\usepackage{color}
\usepackage{float}
\usepackage{xspace}
\usepackage{xcolor}

\usepackage[english]{babel}
\usepackage{colortbl}
\usepackage{youngtab}
\usepackage{ytableau}
\usepackage{amsthm}

\graphicspath{{./graphics/}}

\begin{document}


\title{Fundamental bounds on many-body spin cluster intensities}

\author{Christian Bengs$^{*}$}
\thanks{$^{*}$These two authors contributed equally}
\email{cbengs@berkeley.edu}
\altaffiliation[Also at ]{Chemical Sciences Division, Lawrence Berkeley National Laboratory, Berkeley, CA 94720}
\author{Chongwei Zhang$^{*}$}%
\thanks{$^{*}$These two authors contributed equally}
 \email{zcw1@berkeley.edu}
\author{Ashok Ajoy$^{\dagger}$}%
\thanks{$^{\dagger}$Corresponding author}
\email{ashokaj@berkeley.edu}
\altaffiliation[Also at ]{Chemical Sciences Division, Lawrence Berkeley National Laboratory, Berkeley, CA 94720}
\affiliation{%
 Department of Chemistry, University of California, Berkeley, CA 94720
}%


\date{\today}

\begin{abstract}
Multiple-quantum coherence (MQC) spectroscopy is a powerful technique for probing spin clusters, offering insights into diverse materials and quantum many-body systems. However, prior experiments have revealed a rapid decay in MQC intensities as the coherence order increases, restricting observable cluster sizes to the square root of the total system size. In this work, we establish fundamental bounds on observable MQC intensities in the thermodynamic limit outside the weak polarisation limit. We identify a sharp transition point in the observable MQC intensities as the coherence order grows. This transition points fragments the state space into two components consisting of observable and unobservable spin clusters. Notably, we find that this transition point is directly proportional to the size $N$ and polarization $p$ of the system, suggesting that the aforementioned square root limitation can be overcome through hyperpolarization techniques. Our results provide important experimental guidelines for the observation of large spin cluster phenomena.
\end{abstract}

\maketitle


\section{Introduction}

Nuclear Magnetic Resonance (NMR) techniques serve as exceptional probes of both the macroscopic and microscopic worlds, offering a diverse range of powerful spectroscopic tools for in-depth analysis. Among these, multiple-quantum coherence (MQC) spectroscopy~\cite{drobny_fourier_1978} stands out as a particularly powerful tool, allowing dynamical observation of spin cluster formations~\cite{munowitz_multiple-quantum_1986,baum_multiple-quantum_1986}. As a result, MQC spectroscopy has a broad range of applications across various fields, such as the characterisation of amorphous materials, polymer dynamics, protein organisation and crystal growth~\cite{gleason_hydrogen_1987,petrich_structure_1987,went_magnetic_1992,antzutkin_high-order_1999,antzutkin_multiple_2000,saalwachter_h1_2003,saalwachter_chain_2005,chasse_precise_2011,fayon_evidence_2013,teymoori_multiple-quantum_2013,nowotarski_dynamic_2024}. 

Recently, MQC spectroscopy has also emerged as a valuable tool for the characterisation of quantum many-body dynamics. On the one hand, MQC intensities have been shown to serve as entanglement witnesses~\cite{garttner_relating_2018,wei_exploring_2018}, while on the other hand, MQC intensities may serve as reporters of operator growth~\cite{ramanathan_encoding_2003,cho_decay_2006,alvarez_localization-delocalization_2015,parker_universal_2019,wei_verifying_2020,sanchez_perturbation_2020,dominguez_dynamics_2021,dominguez_decoherence_2021,sanchez_emergent_2022,schuster_operator_2023,li_emergent_2024}. Loosely speaking, a multiple-quantum coherence of order $q$ indicates correlations among at least $q$ spins. MQC intensities thus provide insight into the operator size distribution of the system, which may, in principle, be monitored continuously. Since operator growth is intimately connected to the manifestation of irreversible quantum dynamics, MQC spectroscopy can provide unique insight into universal dynamical behaviour emerging from increasing quantum complexity.

\begin{figure}
\centering
\includegraphics[trim={0 13.5cm 0 0},clip,width=\columnwidth]{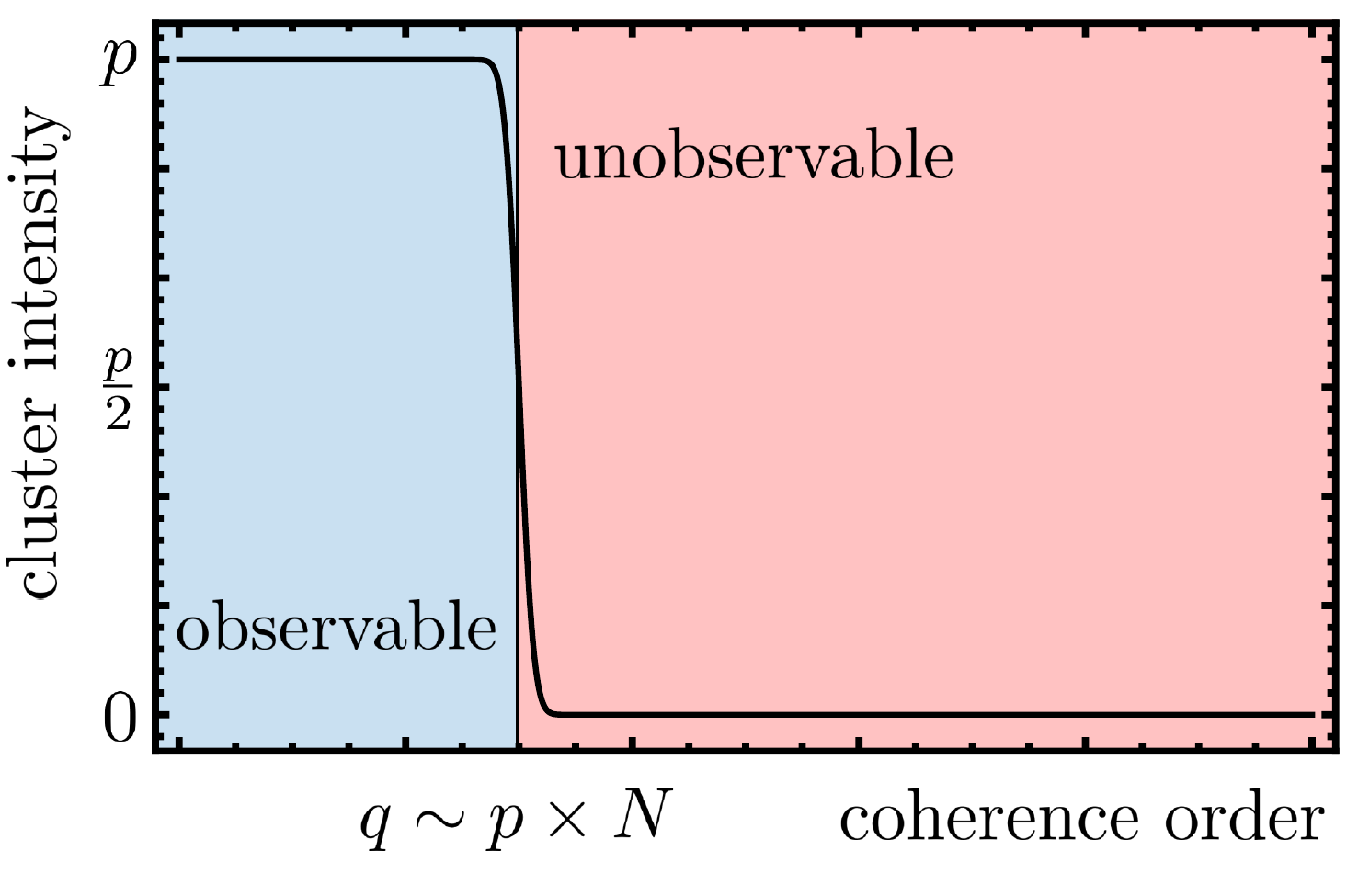}\caption{
\label{fig:cluster_bounds} Observable MQC intensities for a system of $N$ spin-1/2 particles with an initial polarisation $p$. For small coherence order values the observable MQC intensities cling closely to the initial polarisation $p$ of the system. A sharp decay in the observable MQC intensities may be observed for coherence order values falling into a small strip centred at $q\sim p\times N$, with MQC intensities for $q> p\times N$ asymptotically approaching a value of zero. This effective separates the state space into an observable region (blue shaded area) and unobservable region (red shaded area). This behaviour essentially prevents conclusive observations of spin clusters of size $K> p\times N$.}
\end{figure}

The renewed interest in multiple-quantum dynamics raises the question to which extend multiple-quantum NMR spectroscopy may probe large spin cluster effects. Previous results have primarily considered non-selective MQC excitation in the high-temperature limit aiming to build-up large spin clusters in a sequential manner~\cite{baum_multiplequantum_1985-1,munowitz_multiplequantum_1987}. In this case it is well-known that the excited MQC intensities follow a Gaussian distribution with a standard deviation proportional to $\sqrt{N}$, where $N$ is the number of spins within the system. Observation of large MQCs thus becomes increasingly difficult due to a rapid decay of the MQC intensities. The advent of modern hyperpolarisation techniques however has provided unprecedented access to highly non-equilibrium states far outside the high-temperature regime~\cite{maly_dynamic_2008,eichhorn_proton_2014,chen_optical_2015,lilly_thankamony_dynamic_2017,ajoy_orientation-independent_2018,bucher_hyperpolarization-enhanced_2020,eills_spin_2023,dagys_robust_2024}. In combination with selective MQC excitation protocols~\cite{warren_selective_1979,warren_theory_1980,warren_experiments_1981,drobny_selectivity_1997}, which aim to generate many-body states with a well-defined cluster size, hyperpolarisation techniques provide a promising outlook for the observation of large spin cluster effects.

At first, high degrees of polarisation seem to suggest that large spin clusters should become observable. This however turns out to be incorrect. Assuming an ensemble of spin-1/2 particles (pseudo or proper), an initial state of the form $\rho=(\mathbbm{1}/2+p I_{z})^{\otimes N}$ with polarisation $p$ and complete control over the unitary group ${\rm U}(2^{N})$ we derive an idealised MQC intensity distribution. In the thermodynamic limit the MQC intensity distribution may be understood as a convolution of a Gaussian and a uniform distribution with a sharp transition point at $q\sim p \times N$. As schematically shown in figure \ref{fig:cluster_bounds} this separates the state space into two distinct regions of observable and unobservable MQCs. We therefore find that no spin cluster of size $K$ larger than $K\gtrsim p \times N$ may experimentally be observed. Conversely observation of a maximal cluster size $K_{\rm max}$ for a fixed polarisation $p$ may only be possible if the effective spin system size is on the order of $N\sim K_{\rm max}/p$, independent of the excitation scheme. Our results thus provide guidelines on the required polarisation and system size required to observe large spin cluster phenomena. 


\section{Theory}

\subsection{Observable MQC intensities} 

We consider the collective $z$-angular momentum $I_{z}$ of a spin-1/2 ensemble consisting of $N$ spins with eigenstates $\vert i\rangle$ and eigenvalues $\lambda_{i}$. The state  $\rho$ of the system may be expressed as 
\begin{equation}
   \rho = \sum_{q=-N}^{+N}\sum_{\lambda_{i}-\lambda_{j}=q}\rho_{ij}\vert i\rangle \langle j\vert=\sum_{q=-N}^{+N}\rho_{q}.
\end{equation}
The index $q$ is the coherence order of the operator~\cite{ernst_principles_1990}
\begin{equation}
   {[}I_{z},\rho_{q}{]}=q\times \rho_{q},
\end{equation}
with respect to $I_{z}$. Each coherence operator $\rho_{q}$ is a linear combination involving at least $q$ non-trivial shift operators 
\begin{equation}
   \rho_{q}\sim I_{z}^{m}I^{n}_{+}I^{q-n}_{-}.
\end{equation}
Multiple-quantum coherences thus provide useful information on the degree of spin correlations involving $q$ spins or more. 

\begin{figure}
\centering
\includegraphics[trim={0 18cm 0 0},clip,width=\columnwidth]{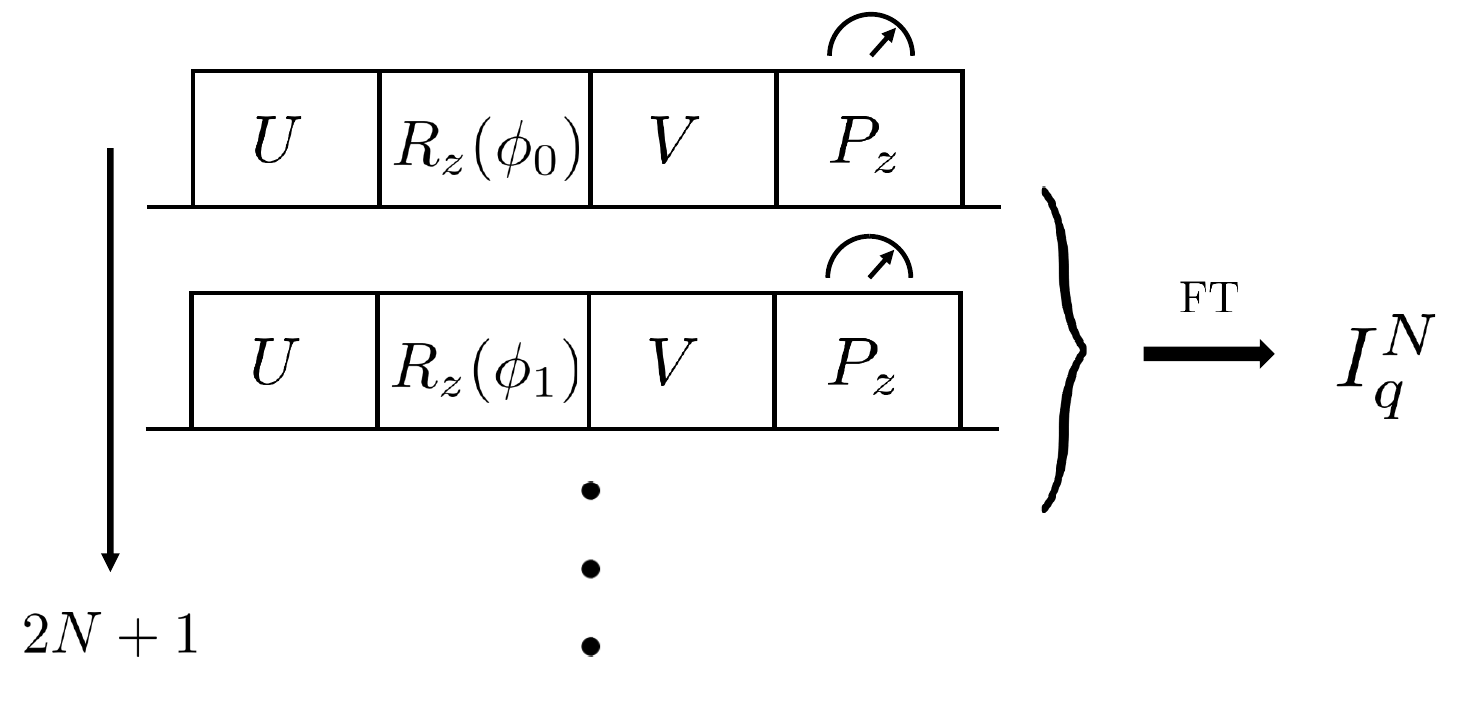}\caption{\label{fig:sel_MQC_exc} Measurement strategy for selective MQC experiments. The system is first subjected to a selective unitary $U_{q}$ maximising coherences within the $q$ coherence subspace followed by phase encoding $R_{z}(\phi_{k})$ for state tomography. For read-out the phase-encoded state is transformed into observable magnetisation $P_{z}$ via $V_{q}$. The unitary $V_{q}$ is chosen to maximise signal components originating from $\rho_{q}$. The protocol is repeated $2N+1$ times varying the phase according to $\phi_{k}=\frac{2\pi k}{2N+1}$. A (real) Fourier transformation of the measurement record isolates the observable MQC intensities $I^{N}_{q}$.
}
\end{figure}

A general MQC experiments typically follows the experimental scheme shown in figure~\ref{fig:sel_MQC_exc}.
Initially, the system is subjected to some unitary $U$ aiming to excite operator components $\rho_{q}$ of the $q$ quantum subspaces. Next, a $z$-rotation achieves phase-encoding of each coherence order subspace
\begin{equation}
R_{z}(\phi_{k})\rho_{q}R_{z}(-\phi_{k})=\rho_{q}e^{i q \phi}, 
\end{equation}
with $\phi_{k}=2\pi/(2N+1)k$. The phase-encoding may be used for partial state tomography, experimentally this may be done via phase cycling~\cite{ernst_principles_1990,levitt_spin_2001}. The observation of the (normalised) collective spin angular momentum
\begin{equation}
 P_{z}=\frac{2}{N}\sum_{j}I_{jz},
\end{equation}
is induced by a second unitary $V$. The normalisation of the observable $P_{z}$ ensures that the MQC intensities remain bounded in the thermodynamic limit. The experiment is repeated for $k\in \{0,\dots, 2N\}$ varying the phase $\phi_{k}$ accordingly. The {\em observable} MQC intensity $I^{N}_{q}$ of coherence order $\vert q\vert $ may then be defined as the real Fourier transform of the measurement record
\begin{equation}
\label{eq:MQC_fourier}
   I^{N}_{q}=\frac{2}{2N+1}\sum_{k=0}^{2N}{\rm Tr}\{P_{z}\hat{V}_{q}\hat{R}_{z}(\phi_{k})\hat{U}_{q}\rho\}\cos(\frac{2\pi kq}{2N+1}).
\end{equation}
Alternatively the Fourier transformation may be represented as the application of a projection operator $\hat{\mathcal{P}}_{\pm q}$ onto the $+q$ and $-q$ quantum coherence subspace
\begin{equation}
   I^{N}_{q}={\rm Tr}\{P_{z} \hat{V}_{q}\hat{\mathcal{P}}_{\pm q}\hat{U}_{q}\rho\}.
\end{equation}

\subsection{Maximal observable cluster intensities} 

The main difference between non-selective and selective MQC experiments is the choice for $U$ and $V$. For non-selective excitation, the forward evolution $U$ is non-specific, leading to excitation of coherences $\{q_{1},q_{2},\dots\}$ with essentially arbitrary amplitudes. The backward evolution is typically taken to be an idealised time-reversal operation $V\simeq U^{\dagger}$. For non-selective excitation in the {\em weak polarisation limit} ($p\ll 1)$ it may be shown that the MQC intensities follow a Gaussian decay~\cite{baum_multiple-quantum_1986,munowitz_multiplequantum_1987}
\begin{equation}
    I^{N}_{q}\propto \exp(-\frac{q^{2}}{N}).
\end{equation}

A selective MQC experiment, on the other hand, focuses on observation of quantum coherences originating from a single coherence subspace. The initial unitary $U=U_{q}$ thus aims to selectively {\em maximise} the coherence operator $\rho_{q}$. Such a unitary may be generated through the concatenation of non-selective excitation elements~\cite{warren_selective_1979,warren_theory_1980}. Furthermore, the backward evolution $V=V_{q}$ may not be related to $U_{q}$ and should aim to maximize the contribution of the signals {\em originating} from $\rho_{q}$. For selective excitation schemes the MQC excitation profile is not known, regardless of the polarisation of the system. 

The conventional Gaussian excitation assumption may thus be broken in at least two ways. The application of highly selective excitation protocols and highly polarised initial states. To account for these scenarios we consider maximal MQC cluster intensity bounds under the following two idealisations: 1) We assume that the initial state of the system displays a uniform local polarisation of degree $-1\leq p\leq +1$
\begin{equation}
    \sigma_{p}=(\frac{1}{2}\mathbbm{1}+p I_{z})^{\otimes N}.
\end{equation}
This choice is motivated by recent progress in hyperpolarisation techniques. 2) We are able to generate any unitary $U\in {\rm U}(2^{N})$ in arbitrary time. The maximal observable cluster intensities $m^{N}_{q}(p)$ may then be formulated as a maximisation problem
\begin{equation}
\label{eq:max_prob}
\begin{aligned}
m^{N}_{q}(p)
&=\max_{U,V \in {\rm U}(2^{N})}{\rm Tr}\{P^{\dagger}_{z} \hat{V}\hat{\mathcal{P}}_{\pm q}\hat{U}\sigma_{p}\}
\\
&=\max_{U,V \in {\rm U}(2^{N})}\sum_{Q_{q}^{k}\in \mathcal{B}_{\pm q}}{\rm Tr}\{P^{\dagger}_{z}\hat{V}Q_{q}^{k}\}{\rm Tr}\{(Q_{q}^{k})^{\dagger}\hat{U}\sigma_{p}\},
\end{aligned}
\end{equation}
where $\mathcal{B}_{\pm q}$ is an orthonormal basis for the $+q$ and $-q$ coherence subspace. An explicit solution to the maximisation problem appears challenging, instead we bound the maximal cluster intensities from above and below. From the definition of $I^{N}_{q}$ (equation~\ref{eq:MQC_fourier}) we have the trivial upper bound $m^{N}_{q}(p)<2p$. We may further bound $m^{N}_{q}(p)$ as follows 
\begin{equation}
\begin{aligned}
m^{N}_{q}(p)<\vert {\rm Tr}\{P^{\dagger}_{z}\hat{V}Q_{q}\}\vert_{\rm max}\vert {\rm Tr}\{Q^{'}_{q}\hat{U}\sigma_{p}\}\vert_{\rm max},
\end{aligned}
\end{equation}
where we maximise the individual factors over $V,U \in {\rm U}(2^{N})$, and two different $Q_{q}, Q^{'}_{q} \in \mathcal{B}_{\pm q}$. As shown in appendix~\ref{app:MQC_proj} the product of the individual factors is maximised by
\begin{equation}
\begin{aligned}
M^{N}_{q}(p)&={\rm Tr}\{P^{\dagger}_{z}\hat{V}Q_{q}\}\vert_{\rm max}\vert {\rm Tr}\{Q^{'}_{q}\hat{U}\sigma_{p}\}\vert_{\rm max}
\\
&=\vert\vert\Lambda^{\downarrow}_{r}(P_{z})\vert\vert_{2}\times 
\vert\vert\Lambda^{\downarrow}_{r}(\sigma_{p})-\Lambda^{\uparrow}_{r}(\sigma_{p})\vert\vert_{2}.
\end{aligned}
\end{equation}
Here, $\Lambda_{r}^{\uparrow}(A)$ denotes a vector of the first $r$ eigenvalues of $A$ when arranged in ascending order, whereas $\Lambda_{r}^{\downarrow}(A)$ denotes a vector of the first $r$ eigenvalues of $A$ when arranged in descending order. The value of $2r=R^{N}_{q}$ on the other hand is chosen according to the largest possible matrix rank $R^{N}_{q}$ among all operators with coherence order $q$ for fixed $N$
\begin{equation}
\begin{aligned}
R^{N}_{q}=2\sum_{j=0}^{q-1}\min\bigg\{
&\sum_{m=0}^{\left\lceil\frac{N-q}{2}\right\rceil}\binom{N}{N-j- 2 m q},
\\
&\sum_{m=0}^{\left\lceil\frac{N-q}{2}\right\rceil}\binom{N}{N-j- 2 m q-q}
\bigg\}.
\end{aligned}
\end{equation}
For details we refer to appendix~\ref{app:MQC_rank}. We may thus bound $m^{N}_{q}(p)$ from above by
\begin{equation}
    B^{N}_{q}(p)\leq\begin{cases}
        2p\hspace{31pt}{\rm for}\quad 2p\leq M^{N}_{q}(p),
        \\
        M^{N}_{q}(p)\quad{\rm for}\quad M^{N}_{q}(p)< 2p.
    \end{cases}
\end{equation}
A lower bound $b^{N}_{q}(p)\leq m^{N}_{q}(p)$ is given by
\begin{equation}
\begin{aligned}
b^{N}_{q}(p)=\Lambda^{\uparrow}_{r}(P_{z})\cdot \Lambda^{\uparrow}_{r}(\sigma_{p})+\Lambda^{\downarrow}_{r}(P_{z})\cdot \Lambda^{\downarrow}_{r}(\sigma_{p}).
\end{aligned}
\end{equation}
This bound is tight and therefore in principle reachable (see appendix~\ref{app:lower_upper_bound}). Again, we refer to the supplementary material for details. The maximal observable cluster intensities are thus bounded by $b^{N}_{q}(p)\leq m^{N}_{q}(p)\leq B^{N}_{q}(p)$.

\section{Cluster intensity bounds} 

\subsection{Explicit cluster intensity bounds} 

\begin{figure}
\centering
\includegraphics[trim={0 5cm 0 0},clip,width=\columnwidth]{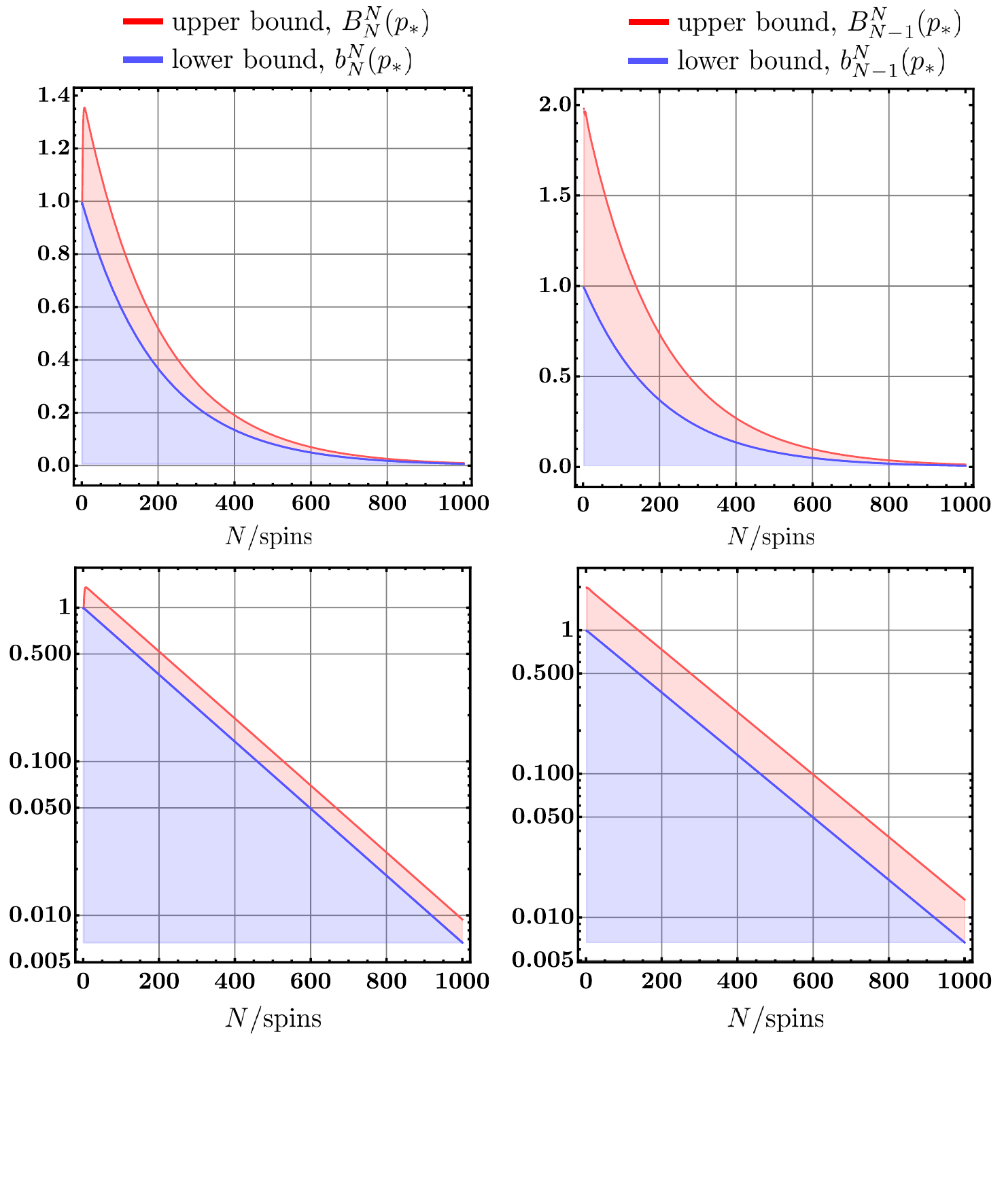}\caption{
\label{fig:cluster_bounds_finitep} Spin cluster intensity bounds as a function of the number of spins $N$ for fixed polarisation $p_{*}=0.99$. Top row: Intensity bounds $b^{N}_{N}(p_{*})$, $B^{N}_{N}(p_{*})$ and $b^{N}_{N-1}(p_{*})$, $B^{N}_{N-1}(p_{*})$ on a linear scale. Bottom row: Intensity bounds $b^{N}_{N}(p_{*})$, $B^{N}_{N}(p_{*})$ and $b^{N}_{N}(p_{*})$, $B^{N}_{N}(p_{*})$ on a semilog scale emphasizing an exponential decay.}
\end{figure}

We first consider some explicit cases which may be solved analytically. For single-quantum coherences ($q=1$) it essentially follows by definition that the maximal rank satisfies $R^{N}_{1}=2^N$. The upper bound $B^{N}_{1}(p)$ and lower bound $b^{N}_{1}(p)$ are then given by 
\begin{equation}
\begin{aligned}
B^{N}_{1}(p)=2p,\; b^{N}_{1}(p)=p,
\end{aligned}
\end{equation}
independent of the size of the system. This result reflects the fact that within a collective measurement framework MQC intensities with $q=1$ cannot distinguish clusters of size $K\geq 1$ or provide any insight into the information scrambling process. In example a large $q=1$ intensity does not exclude the possibility of large information scrambling within the system. For the $q\in \{N-1,N\}$ the maximal matrix ranks are $R^{N}_{N-1}=4$ and $R^{N}_{N}=2$. For the lower intensity bounds we then find
\begin{equation}
\begin{aligned}
b^{N}_{N-1}(p)=\frac{2^{1-n}}{n}(&(1-p)^{n-1}(1+p-n)
\\
&-(1+p)^{n-1}(1-p-n)),
\end{aligned}
\end{equation}
and
\begin{equation}
\begin{aligned}
&b^{N}_{N}(p)=2^{-n}((1+p)^{n}-(1-p)^{n}).
\end{aligned}
\end{equation}
We observe that for $p=1$ the lower bounds equal unity $b^{N}_{N-1}(1)=b^{N}_{N}(1)=1$ regardless of the size of the system. This also holds more generally, in example $b^{N}_{q}(1)=1$. Since these bounds are {\em tight}, the initially pure state $\sigma_{p}$ with $p=1$ enables observation of spin clusters of any size. Consider now the case with an initial polarisation of $p_{*}=0.99$ deviating only 1\% from unity. The upper and lower intensity bounds for $q\in \{N-1,N\}$ with growing system size are shown in figure \ref{fig:cluster_bounds_finitep}. 
The top row shows a rapid decay of both $b^{N}_{N}(p_{*})$, $B^{N}_{N}(p_{*})$ and $b^{N}_{N-1}(p_{*})$, $B^{N}_{N-1}(p_{*})$ as the system size increases. The bottom row shows the bounds on a semilog scale stressing the exponential behaviour of the intensity decay. We may further observe that the upper and lower bounds decay at a similar rate. For large degrees of polarisation we find that the 1/$e$ crossing is approximately given by $N\sim 2/(1-p)$. This implies that observation of large spin clusters on the order of $q\sim N$ would require polarisation levels on the order of $p\gtrsim 1-2/N$. Polarisation levels of this magnitude however are challenging, rendering an observation of these clusters impractical.

\subsection{Large spin cluster intensity bounds} 

\begin{figure}
\centering
\includegraphics[trim={0 3cm 0 0},clip,width=\columnwidth]{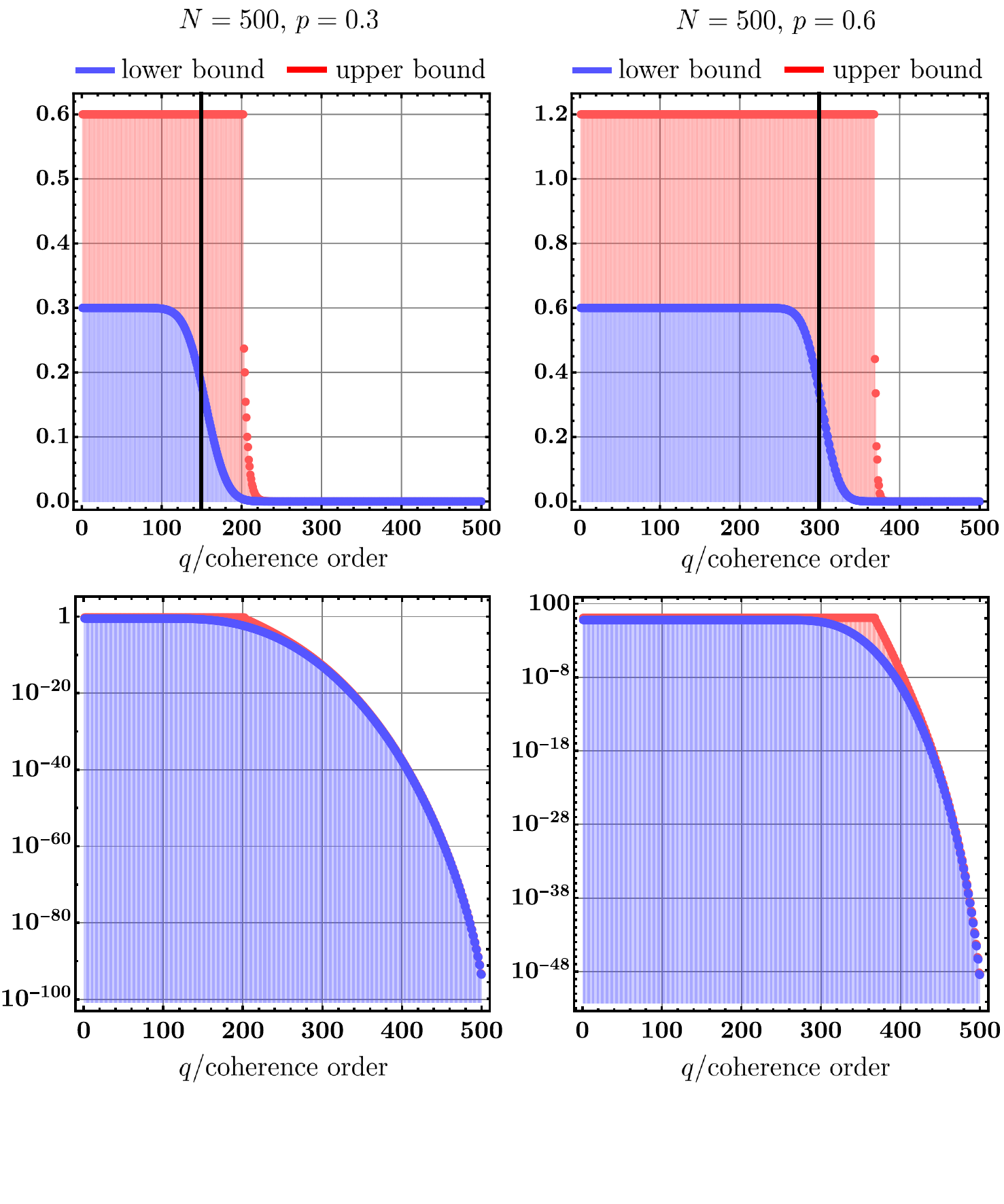}\caption{
\label{fig:cluster_bounds_exact} Upper (red) and lower (blue) observable spin cluster intensity bounds for a fixed system size of $N=500$ and polarisation levels $p\in \{0.3, 0.6\}$ as a function of the coherence order $q$. Top row: Intensity bounds $b^{N}_{q}(p)$, $B^{N}_{q}(p)$ on a linear scale. Bottom row: Intensity bounds $b^{N}_{q}(p)$, $B^{N}_{N}(p)$ on a semilog scale. The upper bounds and lower bounds remain approximately constant at $2p$ and $p$, respectively, until a sharp transition point is reached. For the lower intensity bounds the transition point is indicated by a black line anchored at the point $q\sim p \times N$.}
\end{figure}

The rapid decay of the observable spin cluster intensities shown in figure \ref{fig:cluster_bounds_finitep} raises the question which cluster sizes may realistically be observed for a given initial polarisation $p$ and spin system size $N$. In figure \ref{fig:cluster_bounds_exact} we illustrate the upper and lower cluster bounds for a fixed system size of $N=500$ and polarisation levels $p\in\{0.3, 0.6\}$.
The intensity profiles for other polarisation levels are similar. At first the cluster intensity bounds remain fairly flat and cling closely to $2p$ for the upper bound $B^{N}_{q}(p)$ and $p$ for the lower bound $b^{N}_{q}(p)$. Eventually however the intensity profiles sharply decay to zero over a small range of coherence order values. This is particularly apparent following the intensity bounds on a semilog scale shown at the bottom. The transition point at which the intensities undergo rapid decay depends on the initial polarisation of the system. For higher polarisation values the transition point is shifted to higher coherence order values. This is in agreement with our observation that an increasingly pure initial state enables observation of arbitrarily large spin clusters. We thus find that the upper and lower bounds separate the state space into three distinct regions. An observable region (blue shaded region), an unobservable region (uncoloured region) and a small strip in between which may or may not be observable (red shaded region). The black lines indicate coherence order values for which the lower intensity bound has approximately decayed by half. Outside the weak polarisation limit this crossing point is located at $q\sim p \times N$ as may also be seen from figure~\ref{fig:cluster_bounds_exact} and shown in more detail in appendix~\ref{app:bound_analytics}.

\begin{figure}
\centering
\includegraphics[trim={0 0 0 0},clip,width=\columnwidth]{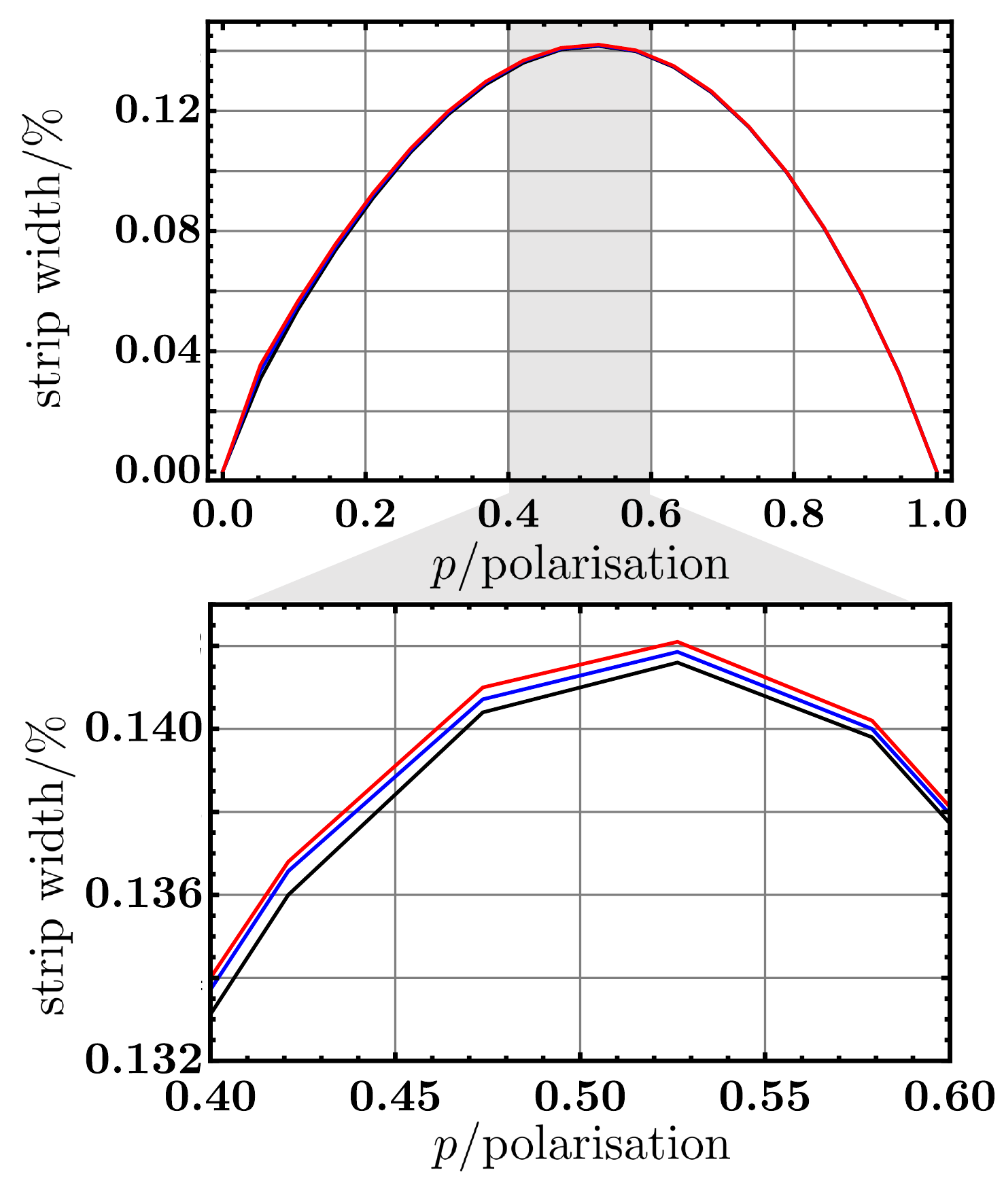}\caption{\label{fig:trans_width} Transition width, defined as the coherence order value difference at which $B^{N}_{q}(p)$ has decayed by half and $b^{N}_{q}(p)$ has decayed by half normalised by the spin system size $N$, as a function of the polarisation of the system. Transitions widths are shown for $N=5000$ (black), $N=7500$ (blue), and $N=10000$ (red). The bottom figure highlights the strip width within the region $p\in {[}0.4,0.6{]}$.}
\end{figure}

For the upper intensity bound we have not been able to derive such a simple relationship. Alternatively we consider the difference between the coherence order value at which $B^{N}_{q}(p)$ has decayed by half and $b^{N}_{q}(p)$ has decayed by half normalised by the spin system size $N$. Since $B^{N}_{q}$ decays rapidly the normalised difference provides a measure of the width of the uncertainty strip. The results are shown figure~\ref{fig:trans_width} for spin systems of size $N=5000$ (black), $N=7500$ (blue) and $N=10000$ (red) as a function of the polarisation of the system. 
In no case does the width of the uncertainty strip exceed $\sim 15\%$ of the total state space. Numerical simulations suggest that the width of the strip converges towards $\sim 14.3\%$ as the system size $N$ increases. We will thus focus on the achievable lower bound ignoring the (relatively) small deviations in cluster observability caused by the uncertainty strip. In the thermodynamic limit we then find that the shape of the lower cluster intensity bounds are well approximated by the convolution of a Gaussian $\mathcal{N}(0,(1-p)N)$ with variance $\sigma^{2}=(1-p)N$ and a uniform distribution $u(-pN,+pN)$ of width $w=2 p N$. 
\begin{equation}
    b^{N}_{q}(p)\propto \mathcal{N}(0,(1-p)N) * u(-pN,+pN),  \quad N\gg 1.
\end{equation}
The width of the transition region for the lower bounds is then approximately given by $~\sim 2\sqrt{6N(1-p)}$ emphasizing a sharp drop in the observable cluster intensities with increasing coherence order. As a consequence the observable region is characterised by $q \lesssim Np-\sqrt{6N(1-p)}$, whereas the unobservable region is characterised by $q \gtrsim Np+\sqrt{6N(1-p)}$. The size of any observable spin cluster $K_{\rm obs}$ in a collective measurement framework thus has to be smaller than 
\begin{equation}
\label{eq:K_obs}
K_{\rm obs}\lesssim Np+\sqrt{6N(1-p)}.
\end{equation}

\section{Experimental implications} 

Let us briefly consider limitations on observable cluster size intensities for the inductive NMR framework. Suppose we carry out an MQC experiment consisting of $2N+1$ scans as shown in figure~\ref{fig:sel_MQC_exc}. The averaged single-shot signal-to-noise ratio (SNR) of the inductively extracted MQC intensities $I^{N}_{q}(p)$ follows~\cite{hoult_signal--noise_1976}
\begin{equation}
\label{eq:SNR_est}
\begin{aligned}
    {\rm SNR}/N &\sim N_{V}V_{\rm coil}\times \frac{I^{N}_{q}(p)}{\sqrt{N} \sigma_{\rm noise}}\sim \frac{\sqrt{N}I^{N}_{q}(p)}{\sigma_{\rm noise}},
\end{aligned}
\end{equation}
where $N_{V}$ represents the spins per unit volume, $V_{\rm coil}$ the coil volume and $\sigma_{\rm noise}$ the standard deviation of the noise. The second relation assumes, for simplicity, that all spins reside within the coil. We observe that while the MQC intensities rapidly decrease with increasing $q$ the averaged SNR is partially recovered with an increasing number of spins ($\propto \sqrt{N}$). Figure~\ref{fig:cluster_bounds_exact} however indicates that the maximal observable spin cluster intensities follow a Gaussian decay beyond $K_{\rm obs}$ strongly suppressing experimental observation of large spin cluster intensities. Suppose we are interested in observing MQC intensities of order $q > N\times p$ with a SNR of at least unity. Equation~\ref{eq:SNR_est} then suggests that the averaged SNR, $\eta$, for a pulse-acquire experiment ($\propto I^{N}_{1}$) needs to be on the order of
\begin{equation}
\begin{aligned}
    \eta\sim \exp\left(\frac{q^2}{N}\right), \quad {\rm for}\;q > N\times p.
\end{aligned}
\end{equation}
This becomes exceedingly difficult to achieve experimentally so that our bound on the observable spin cluster size given by equation~\ref{eq:K_obs} remains valid.

\section{Conclusions}

Within this work we considered fundamental limitations on observable MQC intensities outside the weak polarisation limit. While MQC intensities for weakly polarised systems follow a Gaussian distribution, our results show that MQC intensities bounds for hyperpolarised systems are more appropriately described by a convolution of a Gaussian and a uniform distribution. As a result the MQC intensities undergo a sharp transition at $q\simeq N\times p$. As a consequence of this transition the state space is divided into observable and unobservable multiple-quantum coherences. In particular we find that observable multiple-quantum coherences have to satisfy $q\lesssim Np + \sqrt{6N(1-p)}$. It follows that observation of emergent universal laws, such as size-dependent relaxation phenomena~\cite{dominguez_decoherence_2021,schuster_operator_2023}, becomes increasingly difficult in a collective measurement framework such as NMR. Since we observe these limitations even under ideal unitary evolution, it is plausible that the threshold of observable multiple-quantum coherences could be significantly lower in the presence of pulse errors and decoherence.  

Finally, we would like to emphasize that, while the observable MQC bounds were originally developed within the context of NMR, our conclusions are also applicable in a broader framework. For example, measurement schemes involving generic collective observables of the type $O=\sum_{i} O_{i}$, where the $O_{i}$'s are local operators at sites $i$, and similarly for the initial state of the system. Additionally, our results give further support to the application of state-space restriction methods for the simulation of complex quantum systems~\cite{kuprov_polynomially_2007,dumez_numerical_2010,karabanov_accuracy_2011}. For example we may decompose the simulation trajectory of the collective $x$-magnetisation via 
\begin{equation}
\begin{aligned}
s_{x}(t)&=\sum_{q}(P_{x}\vert \hat{U} \vert \sigma_{p})
\\
&=\sum_{q}(P_{z}\vert \hat{V}\hat{\mathcal{P}}_{\pm q}\hat{V}^{\dagger} \hat{U} \vert \sigma_{p})\leq \sum_{q}m^{N}_{q}(p),
\end{aligned}
\end{equation}
enabling an estimation of the contributions from individual $q$ quantum coherence subspaces, potentially aiding the design of novel polynomially scaling simulation algorithms.

\begin{acknowledgments}
We thankfully acknowledge discussions with the late A. Pines, C. Ramanathan and T. O'Brien, and funding from the ONR (N00014-20-1-2806), CIFAR Azrieli Foundation (GS23-013), and Google LLC. 
\end{acknowledgments}

\section*{Author Declarations}

\subsection*{Conflict of interest}

The authors have no conflicts to disclose.

\section*{Data Availability Statement}
Supporting Mathematica notebooks are available from the authors upon reasonable request.

\appendix

\section{Spectra and matrix ranks of MQC operators}
\label{app:MQC_rank}

For simplicity, we restrict ourselves to the case where the number of spins $N$ and coherence order $q$ are both even, the odd case may be handled in similar fashion. The MQC operators have the following properties:
\\\\
{\bf 1)} For fixed $N$ the maximum achievable matrix rank $R^{N}_{q}$ for a coherence operator of order $q$ is given by
\begin{equation}
R^{N}_{q} = 2\sum_{k=-q/2+1}^{q/2} \min 
\left\{
\sum_{j \text{ odd}}^{\left|jq+k\right|\leq N/2}g_{jq+k},
\sum_{j \text{ even}}^{\left|jq+k\right|\leq N/2}g_{jq+k}
\right\},
\label{eq:rank_opt}
\end{equation}
where $g_{n}=\binom{N}{N/2+n}$ is the degeneracy of the eigenvalue $M_{z}=n$. 
\\
{\bf 2)} We can always find $\frac{1}{2}\times R_{q}^{N}$ linear independent commuting hermitian operators in the subspace of order $q$. 
\\
\noindent 
{\bf 3)} The eigenvalues of any $q$ order MQC operator are of the form $\left\{\pm\lambda_{1}, \pm\lambda_{2},\dots, \pm\lambda_{R_{q}^{N}/2}, 0,\dots, 0 \right\}$, where the eigenvalues $\lambda_{i}$ can be any non-negative real number.
\\\\
\noindent \textit{Proof:} 

{\bf 1)} The proof can be done graphically. Consider a general $q$ order operator $O_{q}$, which may connect different Zeeman manifolds. For example, take $N=4$ and $q=2$, this operator can be broken into two Zeeman 'chains' connecting different Zeeman manifolds: $-2\rightarrow 0 \rightarrow 2$ and $-1\rightarrow 1$. For general $q$, $O_{q}$ can be broken into $q$ chains connecting the manifolds $M_{z}=jq+k$, in which $\left|k\right|\leq q/2$ and $j$ is integer. The rank of the operator $O_{q}$ will then be the sum of the rank of each chain. 
\begin{figure}[h]
    \centering
    \includegraphics[width=\linewidth]{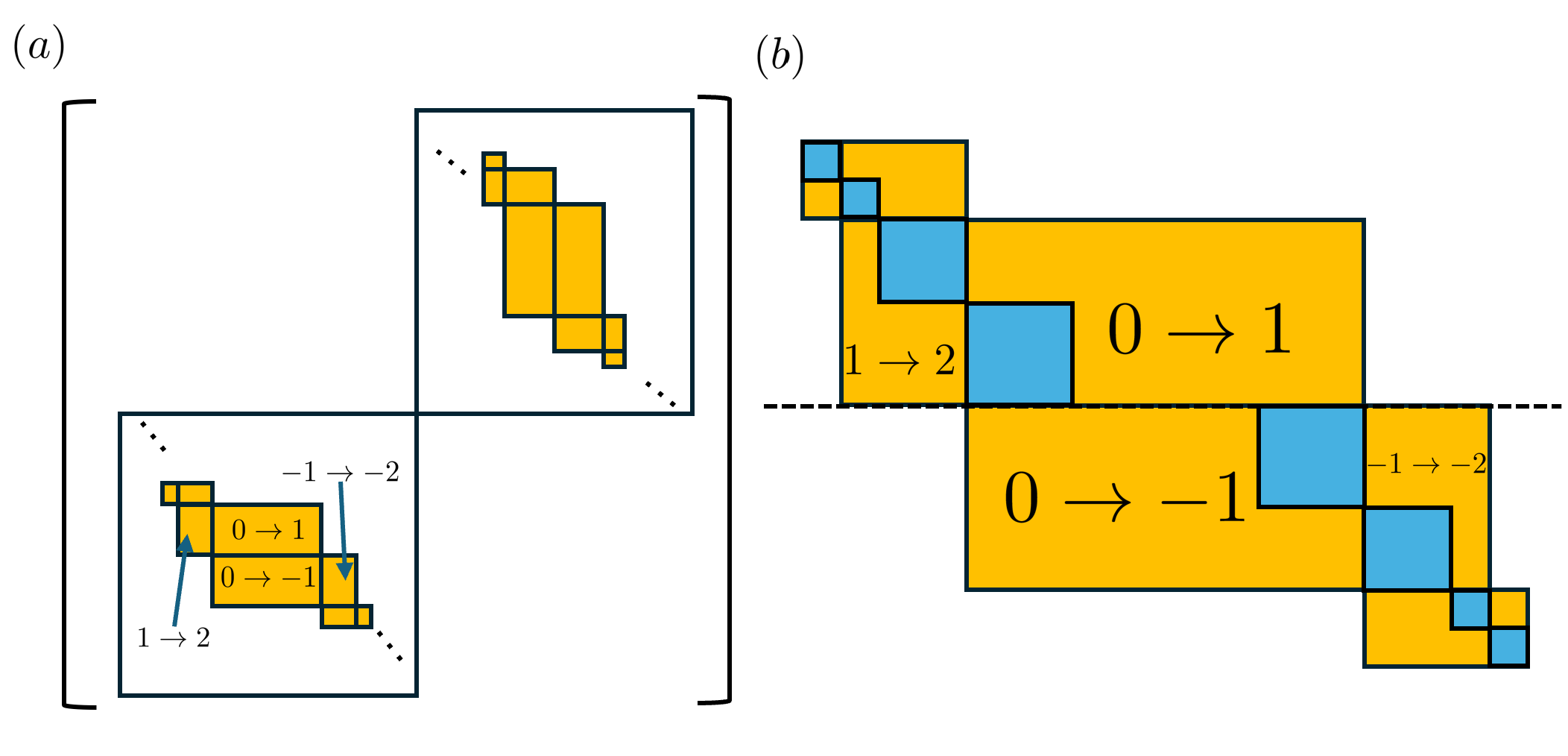}
    \caption{a) Matrix representation of the operator $O_{q}$ upon row and column permutations so that individual Zeeman chains are grouped together. The resulting zigzag patterns across the diagonal are connected via complex conjugation. Arrows such as $0\to 1$ refer to the components connecting  the $M_{z}=0\cdot q+k$ and $M_{z}=1\cdot q+k$ manifold. b) Detailed representation of the lower diagonal zigzag shown in a). Blue block matrices indicate the identity, whereas orange shaded parts are taken to be zero.}

    \label{fig:rank}
\end{figure}
To extract the rank of one of the chains, we permute rows and columns of this chain. The resulting submatrices form a pair of zigzag patterns that are adjoint to each other, as shown in figure \ref{fig:rank} (a). As a result, these two zigzags have the same rank. To get the rank of the rank of zigzag, first we notice that the rank of any one of the two zigzags will be less than the number of rows and number of columns:

\begin{equation}
\begin{aligned}
    \rm{rank(zigzag)}
    &\leq 
    \min 
    \left\{
    \rm{rows(zigzag)},
    \rm{columns(zigzag)}
    \right\}\\
    &=
    \min 
    \left\{
    \sum_{j \text{ odd}}^{\left|jq+k\right|\leq N/2}g_{jq+k},
    \sum_{j \text{ even}}^{\left|jq+k\right|\leq N/2}g_{jq+k}
    \right\}.
    \label{eq:rank}
\end{aligned}
\end{equation}
To show that equality may be achieved we construct a specific zigzag that saturates the inequality. Without loss of generality we consider the case where the number of rows in the bottom-left zigzag is greater than the number of columns. As shown in figure \ref{fig:rank} (b), in each submatrix, part of it is taken to be identity matrix (blue) while the other parts may be taken to 0 (orange). It is not difficult to see that such an arrangement may always be achieved. The rank of this zigzag is thus equal to the accumulated dimension of all the identity matrix, which equals the number of rows the zigzag. Summing now over all pairs of zigzags leads to equation~\ref{eq:rank}.


{\bf 2)} This result follows easily from {\bf 1)}. Take for example a zigzag matrix constructed as described above. If we take one non-zero element in the bottom-left zigzag and the corresponding element in upper-right one, we get $R_{N}^{q}/2$ linear independent and commuting hermitian operators, each with eigenvalue $\pm 1$.

{\bf 3)} From the rotational symmetries of the MQC operators we have 
\begin{equation}
R_{z}\left(\pi/q\right)O_{q}R_{z}^{\dagger}\left(\pi/q\right)=-O_{q}.
\end{equation}
Suppose now $\psi$ is an eigenstate of $O_{q}$, we then have
\begin{equation}
    O_{q}\left(R_{z}\left(\pi/q\right)\left|\psi\right\rangle\right)=R_{z}\left(\pi/q\right)O_{q}\left|\psi\right\rangle=-\lambda\left(R_{z}\left(\pi/q\right)\left|\psi\right\rangle\right),
\end{equation}
so that both $\pm\vert \lambda\vert$ are eigenvalues of $O_{q}$. The spectrum thus takes the form
\begin{equation}
    {\rm spec}(O_{q})=\left\{\pm\lambda_{1}, \pm\lambda_{2},\dots, \pm\lambda_{R_{q}^{N}/2}, 0,\dots, 0 \right\}.
\end{equation}
 An example of $O_{q}$ with this set of eigenvalue is a linear combination of the $R_{N}^{q}/2$ linear independent commuting operators constructed in ${\bf 2)}$, with weights $\{\lambda_{1},...,\lambda_{R_{q}^{N}/2}\}$.

Going through similar arguments we find that on the more general case the maximal rank may be evaluated according to the concise formula
\begin{equation}
\begin{aligned}
R^{N}_{q}=2\sum_{j=0}^{q-1}\min\bigg\{
&\sum_{m=0}^{\left\lceil\frac{N-q}{2}\right\rceil}\binom{N}{N-j- 2 m q},
\\
&\sum_{m=0}^{\left\lceil\frac{N-q}{2}\right\rceil}\binom{N}{N-j- 2 m q-q}
\bigg\}.
\end{aligned}
\end{equation}

\section{Cluster intensity bounds}

\subsection{Projections into the $q$ order subspace}
\label{app:MQC_proj}

Suppose we are given two hermitian operator $A$, $B$ and we are aiming to maximise their overlap $\left(A|B\right)=\mathrm{Tr}\left(A^{\dagger}B\right)$ under the application of some unitary $U$. Given full control over the unitary group the maximum overlap is achieved by optimal alignment of the eigenvalues of $A$ and $B$~\cite{sorensen_universal_1990}
\begin{equation}
    \max_{U\in {\rm U}(2^{N})}\left(B\right|\hat{U}\left|A\right)=\sum_{i=1}^{d_{N}}
    \lambda^{\uparrow}_{i}\left(A\right)\lambda^{\uparrow}_{i}\left(B\right).
\label{eq:overlap_1}
\end{equation}
Here, $\lambda^{\uparrow}_{i}\left(A\right)$, $\lambda^{\uparrow}_{i}\left(B\right)$ are the eigenvalues of $A$, $B$ arranged in decreasing order, and $d_{N}$ is the dimension of the Hilbert space. Consider now the case in which $B_{q}$ represents a {\em normalised} hermitian MQC operator of order $q$ 
\begin{equation}
    (B_{q}\vert B_{q})=1.
\end{equation}
Since the eigenvalue of any MQC operator come in pairs we then find
\begin{equation}
\begin{aligned}
    \max_{U\in {\rm U}(2^{N})}\left(B_{q}\right|\hat{U}\left|A\right)
    &=
    \sum_{i=1}^{d_{N}}
    \lambda^{\uparrow}_{i}\left(A\right)\lambda^{\uparrow}_{i}\left(B_{q}\right)\\    
    &=
    \sum_{i=1}^{r}
    \{\lambda^{\uparrow}_{i}(A)-\lambda^{\uparrow}_{d_{N}+1-i}(A)\}
    \lambda^{\uparrow}_{i}(B_{q})\\
    &\leq
    \sqrt{
    \sum_{i=1}^{r}
    \{\lambda^{\uparrow}_{i}\left(A\right)-\lambda^{\uparrow}_{d_{N}+1-i}\left(A\right)\}^{2}/2
    }\\
    &=\left\|
    \Lambda^{\uparrow}_{r}
    \left(A\right)- \Lambda^{\downarrow}_{r}
    \left(A\right)
    \right\|_{2}/\sqrt{2}.
\end{aligned}
\label{eq:overlap_2}
\end{equation}
Here, $2r$ represents the rank of $B_{q}$, $\Lambda_{r}^{\uparrow}(A)$ denotes a vector of the first $r$ eigenvalues of $A$ when arranged in ascending order, whereas $\Lambda_{r}^{\downarrow}(A)$ denotes a vector of the first $r$ eigenvalues of $A$ when arranged in descending order. 

The argument above holds for any {\em fixed} $B_{q}$. Suppose now however that we further want to maximise the overlap of $\hat{U}\vert A)$ among all the possible $B_{q}$. According to equation~\ref{eq:overlap_2} we should choose an operator $B_{q}$ with maximal rank $2r=R^{N}_{q}$ (see equation~\ref{eq:rank_opt}). By virtue of the Cauchy-Schwarz inequality we should further choose the weights $\lambda_{i}(B_{q})$ according to
\begin{equation}
\label{eq:weights}
\lambda_{i}\left(B\right)\propto \lambda_{i}\left(A\right)-\lambda_{d_{N}+1-i}\left(A\right)
\end{equation}
to saturate the inequality. This is always possible by property {\bf 3)} of the MQC operators. 

\subsection{Upper and lower intensity bounds}
\label{app:lower_upper_bound}

The signal originating from the $q$ quantum subspace is given by
\begin{equation}
S^{N}_{q}\left(\hat{U},\hat{V}\right)=\left(O_{\mathbf{n}}\right|\hat{V}\hat{\mathcal{P}}_{\pm q}\hat{U}\left|\sigma_{p}\right),
\end{equation}
with $\sigma_{p}=(\mathbbm{1}/2+p I_{z})^{\otimes N}$, where $p$ represent the polarisation of the system and $O_{\mathbf{n}}=\frac{2}{N}\sum_{i}O_{\mathbf{n}i}$ is a collective measurement operator along the direction ${\bf n}$, in example $O_{\mathbf{n}i}=\boldsymbol{\sigma}\cdot \mathbf{n}/2$. The maximal observable cluster intensity $m^{N}_{q}(p)$ can be formulated as a maximisation problem
\begin{equation}
\begin{aligned}
m^{N}_{q}(p)
&=\max_{U,V \in {\rm U}(2^{N})}\left(O_{\mathbf{n}}\right|
\hat{V}\hat{\mathcal{P}}_{\pm q}\hat{U}\left|\sigma_{p}\right)
\\
&=\max_{U,V \in {\rm U}(2^{N})}\sum_{Q_{q}^{k}\in \mathcal{B}_{\pm q}}\left(O_{\mathbf{n}}\right|\hat{V}\left|Q_{q}^{k}\right)\left(Q_{q}^{k}\right|\hat{U}\left|\sigma_{p}\right),
\end{aligned}
\end{equation}
where $\mathcal{B}_{\pm q}$ represents an orthonormal basis for the $q$ and $-q$ coherence subspace. While an explicit solution to the problem is challenging, we can bound the maximum cluster intensity from above and below
\begin{equation}
    b^{N}_{q}(p)\leq m^{N}_{q}(p)\leq B^{N}_{q}(p).
\end{equation}
A trivial upper bound for $m^{N}_{q}(p)$ is given by $m^{N}_{q}(p)\leq 2p$. This may be seen as follows. Utilizing the property of MQC operator, the projection operator $\hat{\mathcal{P}}_{\pm q}$ can be written as a Fourier sum
\begin{equation}
    \hat{\mathcal{P}}_{\pm q}=\frac{2}{2N+1}\sum_{k=0}^{2N}\cos{\left(\frac{2\pi kq}{2N+1}\right)}\hat{R}_{\mathbf{n}}\left(\frac{2\pi kq}{2N+1}\right),
    \label{eq:projection}
\end{equation}
where $\hat{R}_{\mathbf{n}}$ is a rotation operator along $\mathbf{n}$ axis. From the definition of $O_{\mathbf{n}}$, it can be shown that for any $O_{\mathbf{n}}$ $\left(O_{\mathbf{n}}|\sigma_{p}\right)\leq p$. We thus have
\begin{equation}
\begin{aligned}
    m^{N}_{q}(p)
    =\max_{U,V \in {\rm U}(2^{N})}\sum_{k=0}^{2N}\frac{2(O_{\mathbf{n}}\vert\hat{V} \hat{R}_{z}\left(\frac{2\pi kq}{2N+1}\right)\hat{U}\vert\sigma_{p})}{2N+1}\leq 2p.
\end{aligned}
\end{equation}
We can further bound $m^{N}_{q}(p)$ as follows
\begin{equation}
\begin{aligned}
    m^{N}_{q}(p)
    &\leq 
    \max_{U \in {\rm U}(2^{N})}\vert\vert( O_{\mathbf{n}}\vert\hat{V}\hat{\mathcal{P}}_{\pm q}\vert\vert
\max_{V \in {\rm U}(2^{N})}\vert\vert \hat{\mathcal{P}}_{\pm q}\hat{U}\left|\sigma_{p}\right)\vert\vert
    \\
    &=
    \max_{\substack{U \in {\rm U}(2^{N})\\ Q_{q}^{k}\in \mathcal{B}_{\pm q}}}( O_{\mathbf{n}}\vert\hat{V}\vert Q_{q}^{k})
\max_{\substack{V \in {\rm U}(2^{N})\\ Q_{q}^{k'}\in \mathcal{B}_{\pm q}}}(Q_{q}^{k'}\vert\hat{U}\vert\sigma_{p}),
\end{aligned}
\end{equation}
where we inserted an additional projection. Making use of equation~\ref{eq:overlap_2} the individual products are maximised as follows
\begin{equation}
\begin{aligned}
    \max_{\substack{U \in {\rm U}(2^{N})\\ Q_{q}^{k}\in \mathcal{B}_{\pm q}}}
   (O_{\mathbf{n}}\vert \hat{V}\vert Q_{q}^{k})    
   &=\vert\vert\Lambda^{\uparrow}_{r}(O_{\mathbf{n}})-\Lambda^{\downarrow}_{r}(O_{\mathbf{n}})\vert\vert_{2}/\sqrt{2},
    \\
    \max_{\substack{V \in {\rm U}(2^{N})\\ Q_{q}^{k'}\in \mathcal{B}_{\pm q}}}
    (Q_{q}^{k'}\vert\hat{U}\vert\sigma_{p})
    &=
    \vert\vert\Lambda^{\uparrow}_{r}(\sigma_{p})-\Lambda^{\downarrow}_{r}(\sigma_{p})\vert\vert_{2}/\sqrt{2}.
\end{aligned}
\end{equation}
We may thus choose $B^{N}_{q}(p)$ as follows
\begin{equation}
\begin{aligned}
&B^{N}_{q}(p)=
\\&\;\;\;\min\{
    2p,
    \frac{\vert\vert\Lambda^{\uparrow}_{r}(O_{\mathbf{n}})-\Lambda^{\downarrow}_{r}(O_{\mathbf{n}})\vert\vert_{2}
    \vert\vert\Lambda^{\uparrow}_{r}(\sigma_{p})-\Lambda^{\downarrow}_{r}(\sigma_{p})\vert\vert_{2}}
    {2}
\}.
\label{eq:higher_bound}
\end{aligned}
\end{equation}
To estimate a lower bound $b^{N}_{q}(p)$, we choose a unitary operator $U_{p}$ that maximises the projection of $\vert\sigma_{p})$ onto the $\mathcal{B}_{\pm q}$ subspace. Let 
\begin{equation}
\vert X_{p})=\frac{\hat{\mathcal{P}}_{\pm q}U_{p}\left|\sigma_{p}\right)}{\vert\vert\hat{\mathcal{P}}_{\pm q}U_{p}\left|\sigma_{p}\right)\vert\vert},
\end{equation}
be the normalised vector of the maximised projection. Although we do not know the precise form of $X_{p}$ by definition its spectrum must satisfy equation~\ref{eq:weights}. We then find
\begin{equation}
    m^{N}_{q}(p)\geq 
    \max_{V \in {\rm U}(2^{N})} 
    (O_{\mathbf{n}}\vert \hat{V}\vert X_{p})(X_{p}\vert \hat{U_{p}}\vert \sigma_{p}),
\end{equation}
since we have chosen $U_{p}$ independently of $V$. As a result the subsequent maximisation over $V$ may serve as the lower bound $b^{N}_{q}(p)$
\begin{equation}
\begin{aligned}
    b^{N}_{q}(p)
    &=
    \max_{V \in {\rm U}(2^{N})} 
    (O_{\mathbf{n}}\vert \hat{V}\vert X_{p})(X_{p}\vert \hat{U_{p}}\vert\sigma_{p})
    \\
    &=
    \Lambda^{\uparrow}_{r}(O_{\mathbf{n}})\cdot \Lambda^{\uparrow}_{r}(\sigma_{p})+\Lambda^{\downarrow}_{r}(O_{\mathbf{n}})\cdot \Lambda^{\downarrow}_{r}(\sigma_{p}).
\end{aligned}
\label{eq:lower_bound}
\end{equation}
By definition we have $b^{N}_{q}(p)\leq B^{N}_{q}(p)$, equality however is achieved if and only if $\Lambda^{\uparrow}_{r}(O_{\mathbf{n}})\propto \Lambda^{\uparrow}_{r}(\sigma_{p})$, which, within the NMR context, may only be achieved within the weak polariation limit $p\ll 1$.

\section{Analytic approximations for cluster intensity bounds}
\label{app:bound_analytics}

To further explore the relationship between the cluster bounds and the polarisation, we are looking for asymptotic approximations for equation \ref{eq:higher_bound} and \ref{eq:lower_bound}. Here, we confine ourselves to $O_{\mathbf{z}}$ instead of a general $O_{\mathbf{n}}$ since they have the same spectrum. The eigenvalues of $O_{\mathbf{z}}$ and $\sigma_{p}$ may be denoted by $\lambda_{n}\left(O_{\mathbf{z}}\right)$ and $\lambda_{n}\left(\sigma_{p}\right)$, respectively, where $n$ is the magnetic quantum number
\begin{equation}
    \begin{aligned}
        \lambda_{n}\left(O_{\mathbf{z}}\right)&=\frac{2}{N}n,
        \\
        \lambda_{n}\left(\sigma_{p}\right)&=\left(1+p\right)^{N/2+n}\left(1-p\right)^{N/2-n}/2^{N}.
    \end{aligned}
\end{equation}
According to equations \ref{eq:higher_bound}, \ref{eq:lower_bound} we should select the $r=R^{N}_{q}/2$ largest and $r$ smallest eigenvalue of $O_{\mathbf{z}}$ and $\sigma_{p}$. In general $r$ is bounded by the accumulated multiplicities  $\sum^{N/2}_{n=k+1}g_{n}\leq r \leq \sum^{N/2}_{n=k}g_{n}$ for some $k$, then we have
\begin{equation}
\begin{aligned}
    &\sqrt{
    \sum_{i=k+1}^{N/2}\left(
     \lambda_{i}\left(\sigma_{p}\right)-\lambda_{-i}\left(\sigma_{p}\right)
    \right)^{2}g_{i}/2
    }
    \\
    &\hspace{10pt}\leq
    \vert\vert
    \Lambda^{\uparrow}_{r}(\sigma_{p})-\Lambda^{\downarrow}_{r}(\sigma_{p})
    \vert\vert_{2}
    \\
    &\hspace{10pt}\leq
    \sqrt{
    \sum_{i=k}^{N/2}\left(
     \lambda_{i}\left(\sigma_{p}\right)-\lambda_{-i}\left(\sigma_{p}\right)
    \right)^{2}g_{i}/2
    \label{eq:norm_ineq1}
    },
\end{aligned}
\end{equation}
\begin{equation}
\begin{aligned}
    &\sqrt{
    \sum_{i=k+1}^{N/2}\left(
     \lambda_{i}\left(O_{\mathbf{z}}\right)-\lambda_{-i}\left(O_{\mathbf{z}}\right)
    \right)^{2}g_{i}/2
    }
    \\
    &\hspace{10pt}\leq
    \vert\vert
    \Lambda^{\uparrow}_{r}(O_{\mathbf{z}})-\Lambda^{\downarrow}_{r}(O_{\mathbf{z}})
    \vert\vert_{2}
    \\
    &\hspace{10pt}\leq
    \sqrt{
    \sum_{i=k}^{N/2}\left(
     \lambda_{i}\left(O_{\mathbf{z}}\right)-\lambda_{-i}\left(O_{\mathbf{z}}\right)
    \right)^{2}g_{i}/2
    }.
\label{eq:norm_ineq2}
\end{aligned}
\end{equation}
We further have
\begin{equation}
\label{eq:estimate1}
\begin{aligned}
    &\ln g_{k+1}\leq 
    \ln \left(\sum^{N/2}_{n=k+1}g_{n}\right)\leq
    \ln r,
    \\
    &\ln r\leq 
    \ln \left(\sum^{N/2}_{n=k}g_{n}\right)\leq
    \ln g_{k}+\ln\left(N/2-k\right).
\end{aligned}    
\end{equation}
Assuming that $N$ is even, we may make use of equation~\ref{eq:rank_opt}, leading to
\begin{equation}
\label{eq:estimate2}
    \ln g_{q/2}\leq 
    \ln r \leq 
    \ln g_{q/2}+\ln\left(N/2-q/2\right).
\end{equation}
For large $N\gg 1$ and $N/2-k> 1$, we further have $\ln g_{k}\gg \ln\left(N/2-k\right)$ and comparison between equations~\ref{eq:estimate1} and \ref{eq:estimate2} implies
\begin{equation}
    k\sim q/2.
    \label{eq:rank_app}
\end{equation}
The eigenvalue differences in equations \ref{eq:norm_ineq1}, \ref{eq:norm_ineq2} may then be estimated as follows
\begin{equation}
\begin{aligned}
    &\ln\left(
    \vert\vert
    \Lambda^{\uparrow}_{r}(\sigma_{p})-\Lambda^{\downarrow}_{r}(\sigma_{p})
    \vert\vert_{2}
    \right)
    \\
    &\quad\sim
    \max_{N/2\geq i \geq q/2}
    \left[
    \ln
    \left(
     \lambda_{i}\left(\sigma_{p}\right)-\lambda_{-i}\left(\sigma_{p}\right)
     \right)
     +
     1/2\ln
     \left(g_{i}\right)
     \right],
     \\ 
      &\ln\left(
    \vert\vert
    \Lambda^{\uparrow}_{r}(O_{\mathbf{z}})-\Lambda^{\downarrow}_{r}(O_{\mathbf{z}})
    \vert\vert_{2}
    \right)
    \\
    &\quad\sim
    \max_{N/2\geq j \geq q/2}
    \left[
    \ln
    \left(
     \lambda_{j}\left(O_{\mathbf{z}}\right)-\lambda_{-j}\left(O_{\mathbf{z}}\right)
     \right)
     +
     1/2\ln
     \left(g_{j}\right)
     \right]
\end{aligned}
\end{equation}
In similar fashion the lower and upper bounds are approximately given by
\begin{equation}
\begin{aligned}
    &\ln\left(B^{N}_{q}(p)\right)\sim
    \min\bigg\{
    \ln\left(2p\right),
    \\
    &\quad\max_{N/2\geq i \geq q/2}
    \left[
    \ln
    \left(
     \lambda_{i}\left(\sigma_{p}\right)-\lambda_{-i}\left(\sigma_{p}\right)
     \right)
     +
     1/2\ln
     \left(g_{i}\right)
     \right]
     +
     \\
    &\quad\max_{N/2\geq j \geq q/2}
    \left[
    \ln
    \left(
     \lambda_{j}\left(O_{\mathbf{z}}\right)-\lambda_{-j}\left(O_{\mathbf{z}}\right)
     \right)
     +
     1/2\ln
     \left(g_{j}\right)
     \right]
     \bigg\},
\end{aligned}
\end{equation}

\begin{equation}
\begin{aligned}
    \ln\left(b^{N}_{q}(p)\right)\sim
    \max_{N/2\geq j \geq q/2}
    \{
    \ln(
    &\lambda_{j}\left(\sigma_{p}\right)\lambda_{j}\left(O_{\mathbf{z}}\right)
    \\
     &+\lambda_{-j}\left(\sigma_{p}\right)\lambda_{-j}\left(O_{\mathbf{z}}\right)
     )
     +
     \ln
     \left(g_{j}\right)
    \}.
\end{aligned}
\end{equation}
From the asymptotic approximation of $b^{N}_{q}(p)$ we may then deduce that the transition point is given by
\begin{equation}
    q_{c}\sim p \times N.
\end{equation}
The transition point of the upper bound does not seem to admit such a simple approximation. However, we find that the transition point is bounded by a simple expression $Q_{c}< \frac{2p}{1+p^{2}}\times N$.

\nocite{*}
\bibliography{MQC_paper}

\end{document}